\documentstyle[aps,prl,preprint]{revtex}

\newcommand{\LP}{\lambda \Phi^4}

\newcommand{\BE}{\begin{equation}}
\newcommand{\EE}{\end{equation}}
\newcommand{\BA}{\begin{eqnarray}}
\newcommand{\EA}{\end{eqnarray}}
\begin{document}

\title{A variational method from the variance of energy}

\author{Fabio Siringo and Luca Marotta}
\address{Dipartimento di Fisica e Astronomia, 
Universit\`a di Catania,\\
INFN Sez. di Catania and INFM UdR di Catania,\\
via S.Sofia 64, 95123 Catania, Italy}
\date{\today}
\maketitle
\begin{abstract}
A variational method is studied based on the minimum of energy variance.
The method is tested on exactly soluble problems in quantum mechanics,
and is shown to be a useful tool whenever the properties of states are more
relevant than the eigenvalues. In quantum field theory the method provides
a consistent second order extension of the gaussian effective potential.
\end{abstract}
\tightenlines 
\section{Introduction}

In 1873 Lord Rayleigh\cite{rayleigh} described a variational method for calculating the
frequencies of mechanical systems. Since then the Rayleigh-Ritz method has
been an important tool for the approximate solution of physical problems.
In quantum mechanics the method has proven very useful, and provides
an upper bound for the ground state energy, as the exact eigenstate of the
hamiltonian $H$  yields the lowest energy expectation value. 

More generally, the
exact eigenstates are known to be stationary points for the expectation value
of $H$.
That is not a special property of $H$: for any real function $f$, the expectation value
of the operator $f(H)$ can be proven to be stationary at the exact eigenstates of $H$.
Thus, when the search of approximate eigenstates of $H$ is the main issue, the expectation
value of any function $f(H)$ can be used as a functional of the trial eigenstate.
In general  the result is different and depends on the choice of the function $f$ unless the trial 
state is the exact eigenstate of $H$.
This dependence is a measure of the accuracy of the approximate eigenstates, and can be
used as a variational method of calculation whenever the description of states is more important
than the determination of the corresponding energies.

In this paper the variance of $H$ is shown to be the natural choice for a measure of the dependence on $f$.
The outcoming variational method, which we call ``Minimal Energy Variance'' (MEV), is not novel,
having been used since 1955  in numerical calculations\cite{bartlett}. 
The method has not been  very popular as the average of the
square $H^2$ is required, and only recently some of its interesting properties have been shown in numerical
Quantum Monte Carlo calculations\cite{malatesta,umrigar}.

We discuss in some detail the properties of  MEV and show that it can be regarded as a useful
complementary tool for the properties of the eigensates
more than a substitute for the usual variational method.
In that respect the MEV turns out to be of interest even for analytical calculations despite the larger
amount of work required for its evaluation which is comparable to a second order perturbative approximation.
We show that MEV is at least as general as the standard variational method, and that it can be relavant
for the variational treatment of quantum field theories like the scalar theory. 

We begin by defining the MEV and discussing its general properties in Sec.II. In fact MEV is
usually described as a numerical tool in the framework of  Quantum Monte Carlo calculations, 
despite its generality. In Sec.III the method is illustrated by comparison of some results for exactly solvable problems
in quantum mechanics (the harmonic oscillator and the hydrogen atom). In Sec.IV MEV is 
shown to be relevant for the variational treatment of  a scalar theory in order to get a consistent second order
extension of the Gaussian Effective Potential.

\section{Definition and properties}

Denoting by $\vert \Psi\rangle$ a generic state in the Hilbert space, the expectation value of $f(H)$ reads
\BE
\langle f\rangle={{\langle\Psi\vert f(H)\vert\Psi\rangle}\over{\langle\Psi\vert\Psi\rangle}}
\EE
and the stationary condition is
\BE
{{\delta\langle f\rangle}\over{\delta\vert\Psi\rangle}}=
{{\langle\Psi\vert f(H)}\over{\langle\Psi\vert\Psi\rangle}}-
{{\langle\Psi\vert f(H)\vert \Psi\rangle}\over{\langle\Psi\vert\Psi\rangle^2}}\langle\Psi\vert=0
\EE
which is satisfied if
\BE
f(H)\vert\Psi\rangle=\langle f\rangle \vert\Psi\rangle
\label{eigengen}
\EE
The solutions of the eigenvalue problem are the stationary points of $\langle f\rangle$. 
Usually (with some special exception\cite{note}) these solutions are the eigenstates of $H$. Thus
whatever is $f$, the stationary points of $\langle f\rangle$ yield approximate eigenstates of $H$.
However the stationary point does depend on the choice of the function $f$ if
the generic trial state $\vert \Psi\rangle$ belongs to a sub-space which does not contain the exact eigenstate.
For instance it is well known that if the trial state $\vert\Psi\rangle$ is not an exact eigenstate then in general
$\langle H^n\rangle\not=\langle H\rangle^n$.
Thus there is no reason why the simple choice $f(H)=H$ should be the best choice. Actually any different choice
for the function $f$ would give a different weighting of the trial state in the expectation value. While the
usual energy variation seems to be reasonable for the approximate evaluation of the ground state energy, that
might not be the best choice for describing the properties of the eigenstates.

We assume that the trial state $\vert\Psi\rangle$ is closer to the exact eigenstate when its
sensitivity to $f$ is lower. In other words we define a distance  $D$ between  the trial state 
and the exact eigenstate according to
\BE
D=\langle f(H) \rangle-f(\langle H\rangle).
\label{distance}
\EE
Thus any choice of the function $f$ would provide a different variational method as we know that
$D=0$ for any exact eigenstate.
The most simple non trivial choice for $f$ is $f(H)=H^2$ and yields the variance 
$\sigma^2=\langle H^2\rangle-\langle H\rangle^2$ as a viable candidate for the distance $D$.

A formal proof that the exact eigenstates of $H$ are stationary points of $\sigma^2$ is trivial:
\BE
\langle\Psi\vert\Psi\rangle{{\delta \sigma^2}\over {\delta \langle\Psi\vert}}=
\left[H^2\vert\Psi\rangle-\langle H^2\rangle\vert\Psi\rangle\right]-
2\langle H\rangle\left[H\vert\Psi\rangle-\langle H\rangle\vert\Psi\rangle\right]
\EE
and the right hand side vanishes if $\vert\Psi\rangle$ is an eigenvector of $H$.
Moreover $\sigma^2\ge 0$, and it vanishes for any exact eigenvector so that
$\sigma^2$ has a minimum at any eigenvector (not just the ground state).
In practice, whenever a trial state is close enough to an eigenstate, the variance $\sigma^2$
is expected to show a local minimum. The value of $\sigma^2$ at the minimum is a measure
of the accuracy of the corresponding approximate eigenstate.
Moreover the minimum of $\sigma^2$ acquires a deeper physical meaning if related to the
dynamical properties of the state. The vanishing of $\sigma$ for eigenstates can be seen as
a consequence of the time delocalization of the stationary states. According to Heisenberg relations
$\Delta t\approx \hbar/\sigma$, a smaller energy variance allows for a longer survival of the approximate
eigenstate. Thus MEV yields approximate eigenstates that best  resemble  the
exact ones in their dynamical evolution.
In that respect MEV seems to be a complementary tool for the properties of the eigensates
more than a substitute for the usual variational method which always gives the best approximation for the
eigenvalues.

\section{Analytical tests}

The method can be tested on exactly solvable problems: the hydrogen atom and the harmonic oscillator.

\subsection{The hydrogen atom}

In atomic units the hamiltonian of the hydrogen atom is
\BE
H=-{1\over 2}\nabla^2-{1\over r}
\label{hydrogen}
\EE
We choose a two-parameter trial state
\BE
\langle r\vert\Psi\rangle=N(1-\alpha r)e^{-\beta r}
\label{trial1}
\EE
This is the exact ground state for $\alpha=0$, $\beta=1$ while it is the first excited state for $\alpha=\beta=0.5$.
The expectation values of $H$ and $H^2$ are easily evaluated
\BE
\langle H\rangle={\left[T_1(\alpha,\beta)-V_1(\alpha,\beta)\right]/\left[2D(\alpha,\beta)\right]}
\label{H}
\EE
\BE
\langle H^2\rangle={{T_2(\alpha,\beta)+V_2(\alpha,\beta)-S(\alpha,\beta)}\over {4D(\alpha,\beta)}}
\label{H2}
\EE
where $T_1=\beta^2(\alpha^2-\alpha\beta+\beta^2)$, $V_1=\beta(3\alpha^2-4\alpha\beta+2\beta^2)$,
$T_2=\beta^4(3\alpha^2+5\alpha\beta+5\beta^2)$, $V_2=4\beta^2(\alpha^2-2\alpha\beta+2\beta^2)$,
$S=2\beta^3(\alpha^2+6\beta^2)$ and $D=(3\alpha^2-3\alpha\beta+\beta^2)$.
The variance $\sigma^2$ has two local minima for $(\alpha, \beta)$ equal to $(0,1)$ and $(0.5,0.5)$, while
the energy $\langle H\rangle$ only has a saddle point at $(0.5,0.5)$. For $\alpha=\beta$ the two methods yield the
same result: a minimum for $\alpha=\beta=0.5$ where the trial state becomes the first excited state.
For $\alpha=-\beta$, the trial state is quite bad: its behaviour for $r\to 0$ is
\BE
\langle r\vert \Psi\rangle \sim N(1-{1\over 2}\beta^2 r^2+{\cal{O}}(r^3))\sim Ne^{-{1\over 2}\beta^2 r^2}
\EE
The variance $\sigma^2$ is quite sensitive to the shape of the wave function, and in this case 
it fails to show any minimum,
while the energy $\langle H\rangle$ still has a minimum for $\beta=1.5$. It is instructive to study the behaviour of
$\sigma^2$ for a constant ratio $\alpha/\beta=k$. The trial state $\vert \Psi\rangle$ 
may get
very close to an exact eigenstate if $k\approx 0$ or $k\approx 1$. In Fig.1 the variance $\sigma^2$ is shown for
several values of $k$.
We observe a cross-over from a pronounced minimum at $\beta=0.5$ (for $k=1$) 
to a pronounced minimum at $\beta=1$ (for $k=0$).
As $k$ gets negative and moves away from $0$, the minimum value of $\sigma^2$ raises and eventually the minimum
disappears as the trial state becomes worse and worse. Conversely as $k$ approaches $0$, a minimum around
$\beta=1$ deepens until $\sigma^2$ vanishes for $k=0$. The minumum at $\beta=0$ is always present as in that
limit the trial wave function becomes a constant which is an exact unbounded eigenstate with a vanishing energy.
Thus at variance with the standard variational method, MEV may be used for approximating 
excited states without having to insert orthogonality conditions: a local minimum appears whenever the 
trial state is close enough to an exact eigenstate. Moreover the sensitivity of $\sigma^2$ discards bad approximations
as the minimum disappears for the worse trial states. Whenever a minimum is present its value is by itself a measure
of the accuracy of the state as $\sigma^2=0$ for the exact eigenstate.

\subsection{The harmonic oscillator}

Other insights on the method come from the study of the simple harmonic oscillator. Let us consider the 
Hamiltonian
\BE
H= {{p^2}\over 2}+{1\over 2}\omega_0^2 x^2
\label{harmonic}
\EE
which describes an oscillator whose frequency is $\omega_0$. Let us denote by $\vert n,\omega_0\rangle$ the
exact eigenstates with energies $E_n=\hbar\omega_0 (n+1/2)$. As a trial state we may take a linear combination
of the lower energy eigenstates of a generic oscillator whose frequency is $\omega$:
\BE
\vert \Psi\rangle= \vert 0,\omega\rangle+\alpha \vert 1,\omega\rangle
\label{trial2}
\EE
where both $\alpha$ and $\omega$ are variational parameters. 
The trial state is the exact ground state of $H$ for $\alpha=0$ and
$\omega=\omega_0$, while it gives the first excited state for $\alpha\to\infty$ and $\omega=\omega_0$.
The calculation of the expectation values of $H$ and $H^2$ is trivial: in units of $\hbar\omega_0/2$ we may express
them as
\BE
\langle H\rangle= \cosh (\ln x) f_3(\alpha)
\label{H_h}
\EE
\BE
\langle H^2\rangle= 3\sinh^2(\ln x) f_5(\alpha) +f_9(\alpha)
\label{H_h2}
\EE
where $x=\omega_0/\omega$ and $f_n(\alpha)=(1+n\alpha^2)/(1+\alpha^2)$ is a smooth increasing
function of  $\alpha$ ranging from $1$ (at $\alpha=0$) to $n$ (for $\alpha\to\infty$).

The variance follows as 
\BE
\sigma^2={1\over{\cosh^2(\ln \alpha)}}+g(\alpha)\sinh^2 (\ln x)
\label{variance}
\EE
where $g(\alpha)=2(3\alpha^4+6\alpha^2+1)/(1+\alpha^2)^2$ is a smooth increasing function of
$\alpha$ ranging from $2$ (at $\alpha=0$) to $6$ (for $\alpha\to \infty$).

First of all we mention that both methods must predict the exact values of $\alpha$ even for $x\not=1$ 
(i.e. $\omega\not=\omega_0$):  in fact at any $\omega$ the states $\vert 0,\omega\rangle$ and $\vert 1,\omega\rangle$
have different symmetry properties, and thus the trial state can be an eigenstate of parity only for $\alpha=0$ (even)
or $\alpha\to\infty$ (odd). Actually we may observe that $\alpha=0$ and $\alpha\to\infty$ are stationary points of 
$\langle H\rangle$ and $\sigma^2$ for any choice of the parameter $x$. This is evident for $\langle H\rangle$ as
in Eq.(\ref{H_h}) the contributions of $\alpha$ and $x$ are in different factors. We always 
get a minimum for $\alpha=0$ (ground state), while the limit $\alpha\to\infty$ is a maximum (first excited state).
Whatever is $\alpha$, $\langle H\rangle$ has a unique stationary point for $x=1$ where the hyperbolic cosine has a
minimum. From Eq.(\ref{variance}) we see that the variance follows the same path: for any choice of $\alpha$ a minimum
occurs at $x=1$ where the hyperbolic sine vanishes. If we set $x\not=1$, then we can explore the dependence of
the variance on $\alpha$. As we move  from $x=1$ (i.e. $\omega=\omega_0$), the trial state gets worse and worse.
We still find two minima at $\alpha=0$ and $\alpha\to\infty$, but the minimum value of the variance increases as
the state gets worse. In Fig.2 the variance is shown for some values of $x$. From Eq.(\ref{variance}) we see that
at the minima $\sigma^2=g(\alpha)\sinh ^2(\ln x)$ so that we get a larger variance $\sigma^2=6\sinh^2(\ln x)$ at the
first excited state ($\alpha\to\infty$), and a smaller variance $\sigma^2=2\sinh^2(\ln x)$ at the ground state ($\alpha=0$).
Thus the trial state is a better approximation for the ground state than it is for the first excited state.

\section{Scalar theory}
In quantum field theory the properties of vacuum are more relevant than its energy which is not finite anyway.
For instance the symmetry breaking mechanism and the mass of the Higgs boson depend on the structure of the true vacuum.
Thus we argue that the use of MEV could give rise to new insights on  the ground state properties of  
relevant field theories like the scalar theory. 
In fact we show that MEV may be used for improving the Gaussian Effective Potential
(GEP), a useful variational tool which has been discussed by several authors since 
1974\cite{kuti,barnes,chang,weinstein,huang,bardeen,stevenson1,stevenson2}.
The GEP has many merits, and has been successfully applied to physical problems ranging from
electroweak symmetry breaking\cite{ibanez-meier} 
and scalar theories\cite{stevenson2}, to superconductivity
in bulk materials\cite{camarda} and films\cite{abreu}.
A second order extension of the gaussian approximation would be desirable for a better understanding of
the symmetry breaking transition. In fact sometimes the GEP is known to predict a first order transition even 
when the phase change should be continuous.
Moreover, the GEP fails to show a minimum for some ranges of parameters.
Attempts to improve the GEP have not been so successful:
the Post Gaussian Effective Potential (PGEP) discussed
by Stancu and Stevenson\cite{stancu} fails to reach a minimum for any finite
value of the variational parameter which is fixed by the vanishing 
of the second derivative\cite{stevenson0}. A way out has been studied by 
Tedesco and Cea\cite{tedesco} who take the variational parameter fixed at the first order value.
In this paper we point out that the minimum of variance would be a viable tool for determining the variational
parameter, and we show that this choice allows a useful second order extension of the GEP.

The GEP can be seen as an improved first order perturbative approximation. Let us decompose 
the hamiltonian  in two parts as $H=H_\Omega+V_\Omega$ where $H_\Omega$ is any solvable hamiltonian
which depends on the parameter $\Omega$, while $V_\Omega=H-H_\Omega$. 
The decomposition itself depends on  the parameter $\Omega$. The ground state of $H_\Omega$ 
satisfies the eigenvalue
equation 
\BE
H_\Omega \vert\Psi_\Omega\rangle=E_\Omega\vert\Psi_\Omega\rangle.
\label{eigenGEP}
\EE
Then the first order perturbative approximation for the lower eigenvalue of $H$ follows
\BE
E=E_\Omega+\langle\Psi_\Omega\vert V_\Omega\vert\Psi_\Omega\rangle
\label{perturb}
\EE
The minimum of $E$ can be found by a variation of the parameter $\Omega$, and at
the minimum point $\Omega=\Omega_0$ we get the best decomposition of $H$
(in the sense that the first order perturbative approximation yields the lower energy).
However $E$ is the expectation value of the full hamiltonian $H$, and the method is
a genuine variational method: the trial state is the eigenstate $\vert \Psi_\Omega\rangle$
which depends on the parameter $\Omega$ according to Eq.(\ref{eigenGEP}).
In the GEP $H_\Omega$ is the hamiltonian of a free scalar field whose mass is $\Omega$, and its ground state 
$\vert\Psi_\Omega\rangle$ is a gaussian functional of fields. 

The PGEP\cite{stancu} is equivalent to the second order perturbative evaluation
of the vacuum ground state energy (effective potential). It arises from the sum of all the second order connected
one-particle irreducible diagrams without external legs. It can be proven to be equivalent to the 
cumulant expansion discussed by Kleinert\cite{kleinert}, and then the second order correction $\delta E^{(2)}$
is basically equivalent to the variance up to a sign
\BE
\delta E^{(2)}=\langle V_\Omega\rangle^2-\langle V^2_\Omega\rangle=
\langle H\rangle^2-\langle H^2\rangle=-\sigma^2
\label{varianceGEP}
\EE
Thus the minimum of the variance is equivalent to the minimum absolute value of the second order correction.
According to the asymptotic convergence of the perturbative expansion we know that a minimum of the
second order correction is equivalent to a minimum of the error that we expect in the first order expansion.
From this point of view the minimum of the variance  singles out the best perturbative expansion.

The explicit expression for the second order effective potential $V^{(2)}$ has been reported 
in Ref.\cite{stancu} as a function of the vacuum expectation value of the field $\langle\phi \rangle=\varphi$ 
for a scalar theory whose action reads 
\BE
S[\phi]=\int d^dx\left[{1\over 2}\phi(x) \left(-\partial^2+m^2\right)\phi(x)+\lambda\phi^4(x)\right]
\label{lagrangian}
\EE
in a $d$-dimensional Euclidean space. 
The second order effective potential is given by
\BE
V^{(2)}=V^{(1)}+\delta E^{(2)}
\EE
where $V^{(1)}$ is the first order GEP
\BE
V^{(1)}=I_1(\Omega)+{1\over 2} m^2\varphi^2+\lambda\varphi^4+
{1\over 2}I_0(\Omega)\left[m^2-\Omega^2+12\lambda\varphi^2+6\lambda I_0(\Omega)\right]
\EE
and the second order correction reads
\BE
\delta E^{(2)}=-\left\{{1\over 8} I^{(2)}(\Omega)\left[m^2-\Omega^2+12\lambda\varphi^2+
12\lambda I_0(\Omega)\right]^2+8\lambda^2\varphi^2 I^{(3)}(\Omega)+
{1\over 2}\lambda^2 I^{(4)} (\Omega)\right\}.
\EE
Here  $I^{(n)}(\Omega)$ and  $I_n(\Omega)$ are the integrals defined according to\cite{stancu}
\BE
I_1(\Omega)={1\over 2}\int {{d^dp}\over{(2\pi)^d}}\ln(p^2+\Omega^2)
\EE
\BE
I_0(\Omega)=\int {{d^dp}\over{(2\pi)^d}}{1\over{p^2+\Omega^2}}
\EE
\BE
I^{(n)} (\Omega)= n! \int {d^dx}\left[G(x)\right]^n
\EE
where $G(x)$ is the free particle Green function
\BE
G(x)=\int {{d^dp}\over{(2\pi)^d}}{{e^{ipx}}\over{(p^2+\Omega^2)}}
\EE

Most of these integrals are diverging and must be regularized.
The search for the minimum of $V^{(2)}$ yields a gap equation for the free field mass $\Omega$.
A numerical analysis of this gap equation shows that there is no minimum for the second order effective potential, 
while the second order correction by itself (the variance) has a pronounced minimum for
a broad range of the parameters.

The second order correction has been evaluated
as a function of the bare parameters $m$ and $\lambda$, and the variational parameter $\Omega$.
An energy cut-off $\Lambda$ has been inserted in order to regularize all the diverging integrals.
In Fig.3 $\delta E^{(2)}$ versus $\Omega$ is reported, for $d=3$ and for the set of parameters
$m^2/\Lambda^2=-0.06$, $\lambda/\Lambda=0.1$, $\varphi/\sqrt{\Lambda}=0.1$. 
According to Eq.(\ref{varianceGEP}) it turns out 
to be negative; moreover, as it is clear from the figure, the second order correction, while owning a maximum when 
its absolute value is minimum (which is the  minimum of  the variance: see the Fig.3 inset ),  is not bounded: 
this explains why the total effective potential  fails to reach a minimum for any choice of the free mass $\Omega$.

The minimum of $\sigma^2$ yields a best value $\Omega=\Omega_0$
for each  value of the shift  $\varphi$. Insertion in
$V^{(2)}$ gives our second order effective potential. 
This should be compared to the PGEP of Ref.\cite{stancu} where
the best $\Omega$ is obtained by  the vanishing of the second derivative of $V^{(2)}$\cite{stevenson0}.

In Fig.4 our second order effective potential is reported (the same $d=3$ and bare parameter values 
as those in Fig.3 were used). For this set the system is close to its transition point. 
For comparison
in the same figure we also show the standard first order GEP, and the PGEP evaluated according to Ref.\cite{stancu}.
For $d=3$ the system may be regarded as a static statistical model for a phase transition in the three-dimensional space
(Ginzburg-Landau action). The predictions of this model can be tested by comparison with the experimental data on
the phase transition of different systems like superfluids and superconductors. 
Unfortunately the simple GEP predicts a first order transition in this case (while the transition is known 
to be continuous). In Fig.5
an enlargement of the $\varphi=0$ area makes these reasonings more evident: the GEP (dotted line) is an increasing 
function up to a maximum (the point
$\varphi=0$ is a local minimum). Actually the phase transition occurs when the true minimum rises more than the local
minimum (first order transition). Our second order potential (solid line), evaluated by MEV, 
predicts a continuous transition (as it should be), with the point at $\varphi=0$ \emph{always} being 
a local maximum in the broken phase.
Thus the method provides a consistent second order extension of the GEP while retaining its variational character.

We conclude that while MEV  has been recently shown to be a useful tool in numerical Quantum
Monte Carlo calculations, its potentialities have not been fully explored yet. 
Whenever the properties of states are more relevant than the eigenvalues, MEV
provides a viable variational method which can be used in analytical and field theory calculations as a
complementary tool.

\begin{figure}

\caption{The variance $\sigma^2$ for approximate eigenstates of the hydrogen atom.
The trial wave function is defined according to Eq.(\ref{trial1})  with $\alpha=k\beta$ and
$k=$ 1.0, 0.8, 0.6, 0.4, 0.2, 0.0, -0.2, -0.4, -0.6, -0.8. The minimum moves from left to right
when $k$ decreases, and disappears at  $k\approx-0.7$}

\caption{The variance $\sigma^2$ for approximate eigenstates of the harmonic oscillator.
The trial state is defined according to Eq.(\ref{trial2}) with $\alpha$ ranging from $\alpha=0$
(ground state) to $\alpha\to\infty$ (first excited state) while $\omega_0/\omega=x$ is taken to be
$x=1$  (lower curve), $x=1.25$ and $x=1.5$ (upper curve). Approaching the exact eigenstates ($x\to 1$)
the minima decrease and eventually vanish.}

\caption {The second order correction (the variance, up to a sign) for the effective potential in $\LP$ scalar theory. The
values for the parameters are: $d=3$, $m^2/\Lambda^2=-0.06$ and $\lambda/\Lambda=0.1$, 
$\varphi/\sqrt{\Lambda}=0.1$.
Note that it is unbounded,so that the total effective potential 
cannot have a minimum whatever the variational parameter $\Omega$ is;
as shown in the inset graph (an enlargement with $\sigma^2$ unit scaled by a factor $10^{4}$), 
however, it has, by itself, 
a pronounced maximum, making MEV a reasonable alternative to PGEP, 
where the vanishing of $V^{(2)}$ second derivative is required.}

\caption{The second order effective potential evaluated by the method of minimum variance (solid line) for
$d=3$, $m^2/\Lambda^2=-0.06$ and $\lambda/\Lambda=0.1$.
For comparison the PGEP 
(dashed line) and the simple first order GEP (dotted line) are reported.
The effective potential is scaled by
a factor $10^{5}$, while the field shift $\varphi$ is in units of 
$\sqrt{\Lambda}$.}

\caption{An enlargement (the effective potential is scaled by a factor $10^{9}$) of the $\varphi=0$ region in Fig.4; 
the GEP (dotted line) 
predicts a first order transition with the point at $\varphi=0$ being a local minimum; 
instead, the MEV (solid line)  predicts 
a continuous transition (as it should be for the superconductivity), 
thus providing a consistent variational second order extension
of the GEP.}
\end{figure}

\begin{thebibliography} {99}
\bibitem{note} If $f$ is not a monotonic function, then its inverse would be multivalued: for two different
energy eigenvalues $E_1$, $E_2$  we might find $f(E_1)=f(E_2)$ and the eigenstates would be 
degenerate for the operator $f(H)$. Thus the solution of Eq.(\ref{eigengen}) could be a linear combination of
different eigenstates of $H$.
\end{thebibliography}
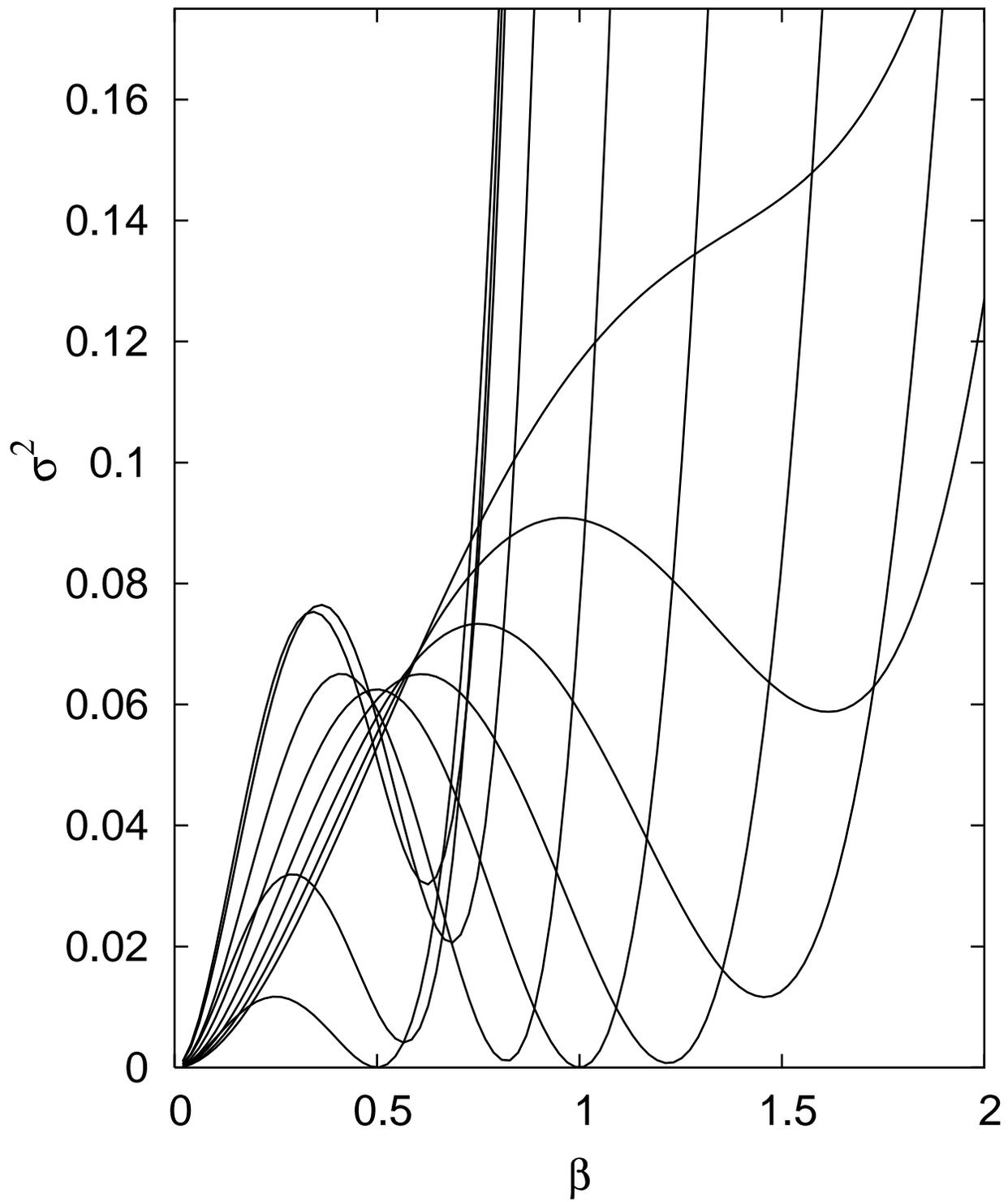
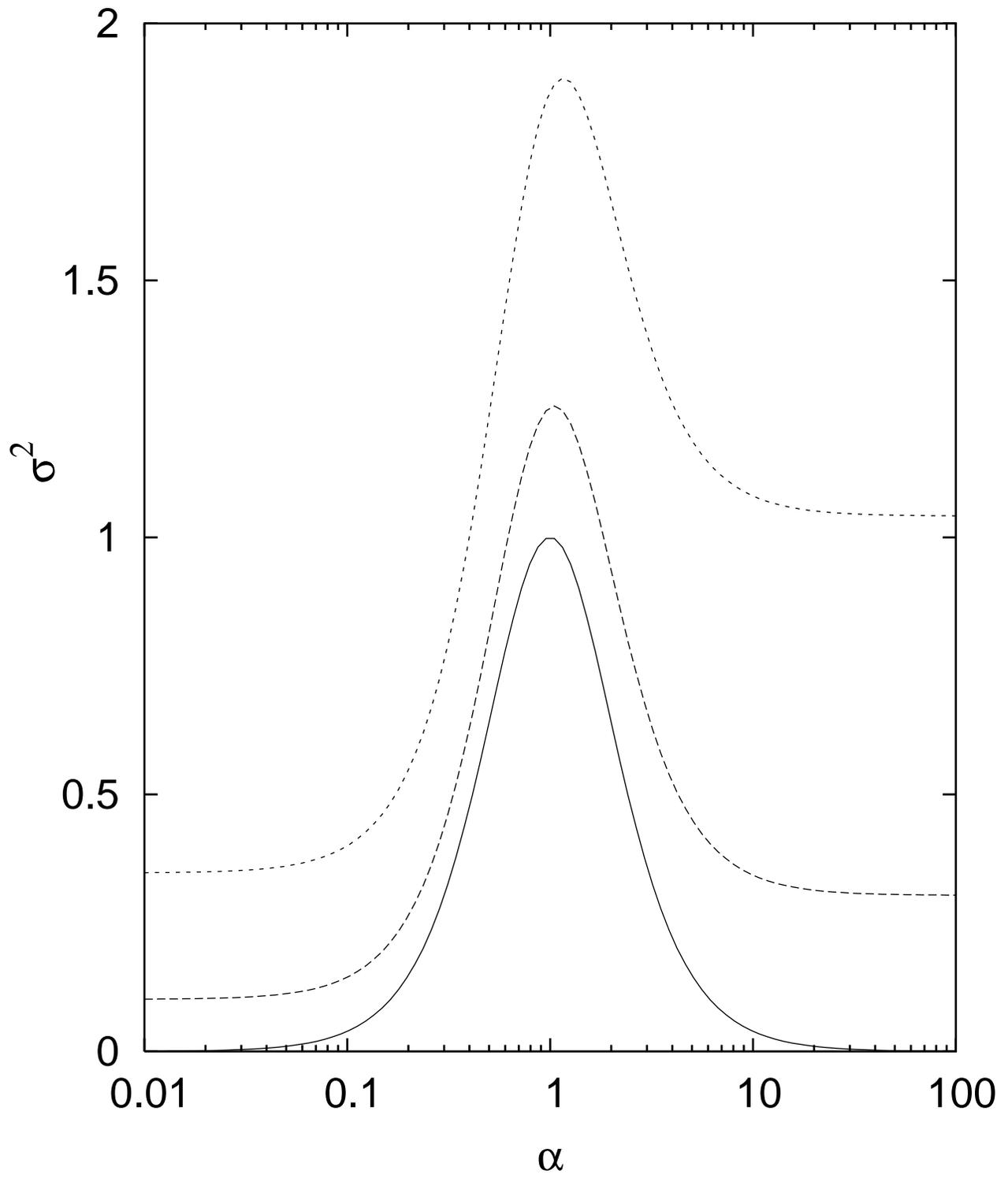
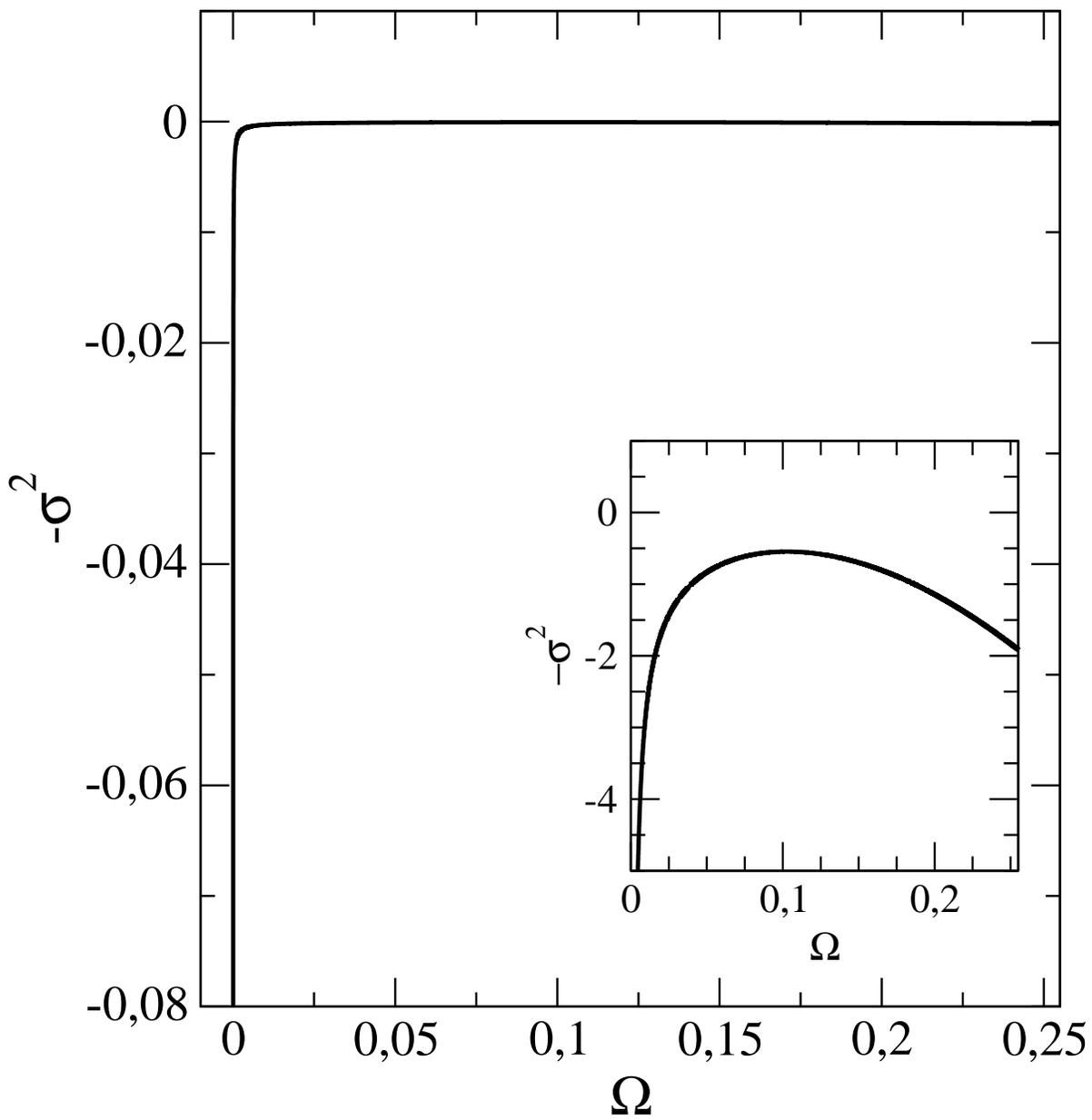
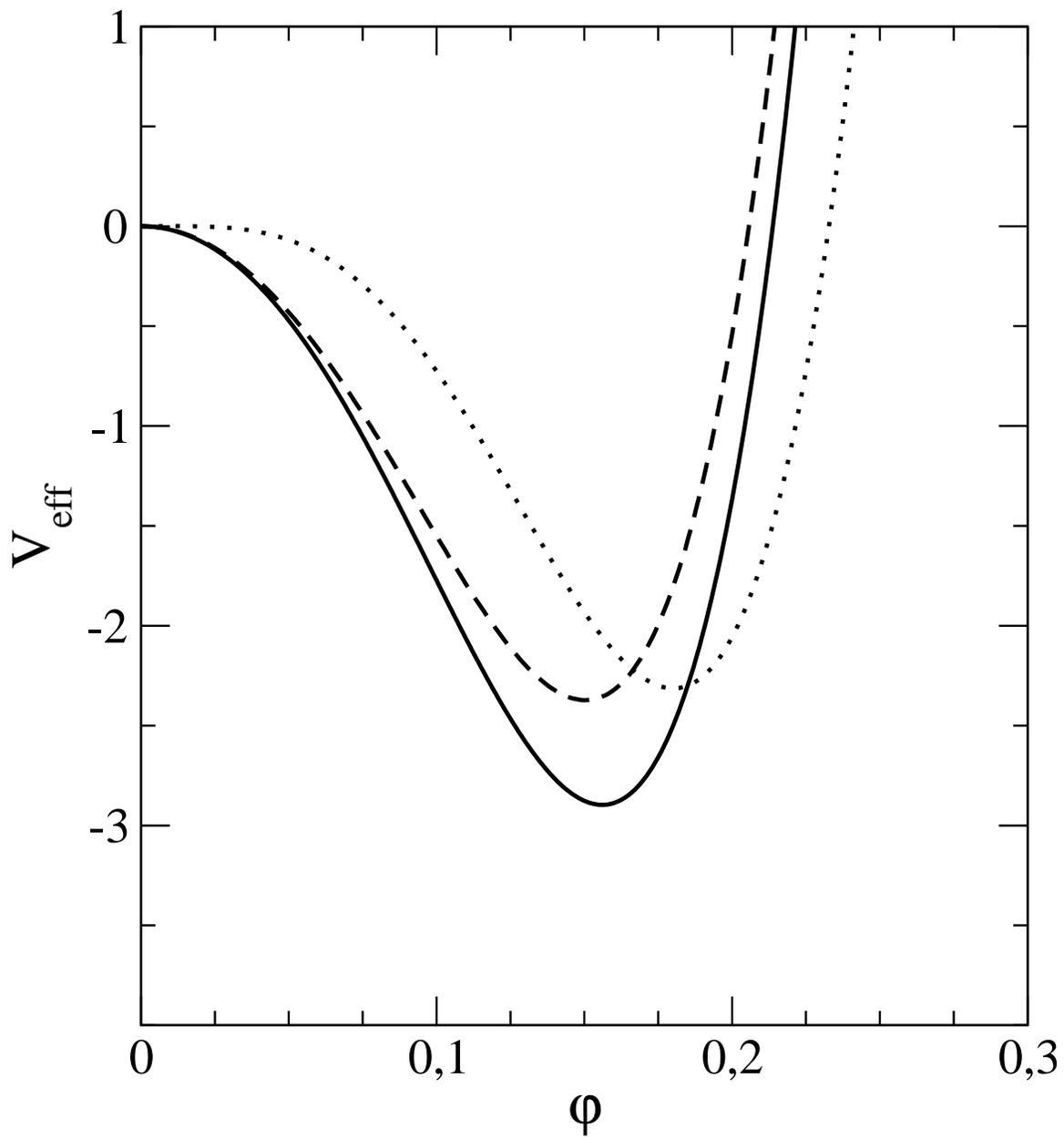
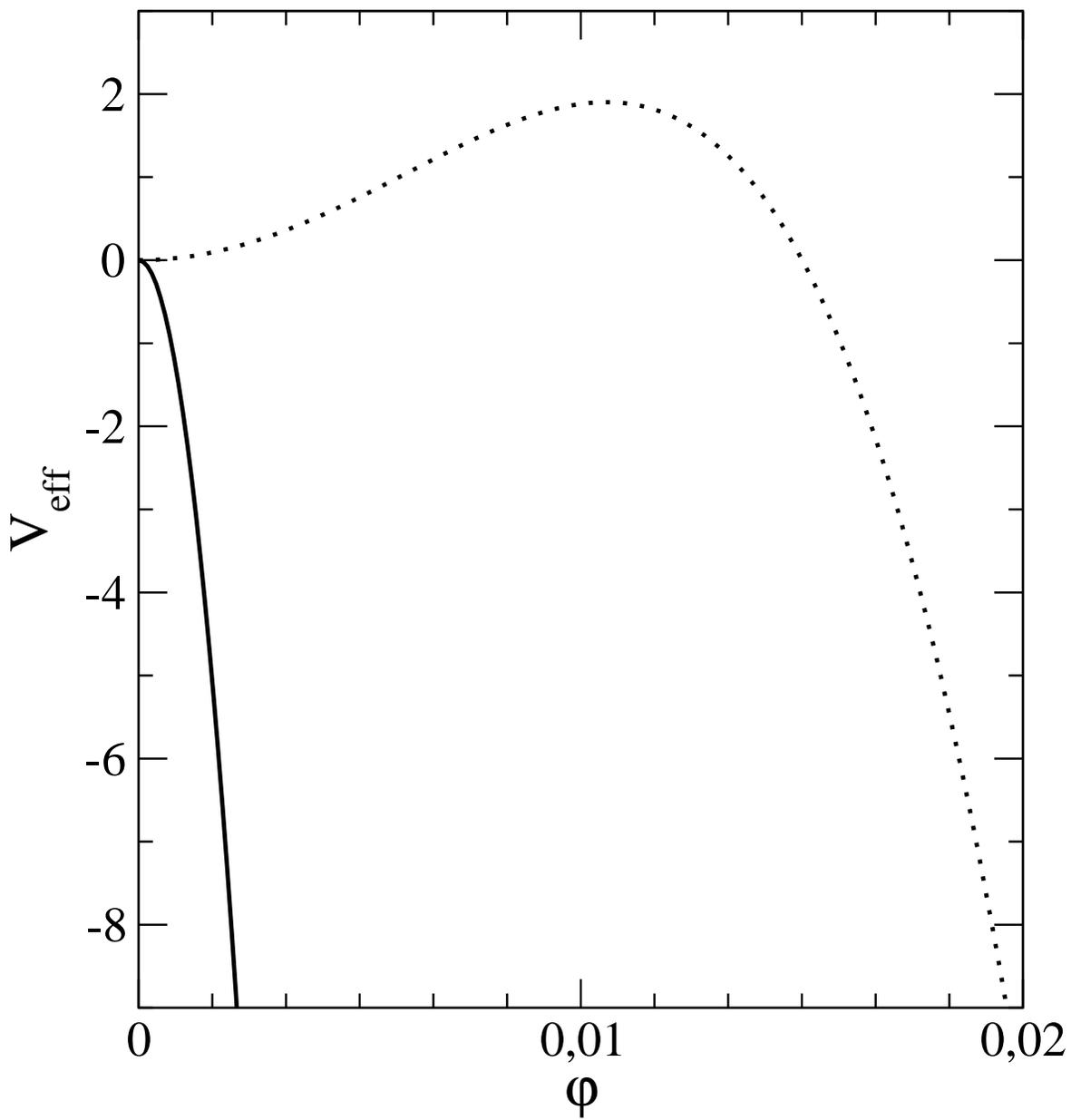
\end{document}